\documentclass[sigplan]{acmart}

\def\BibTeX{{\rm B\kern-.05em{\sc i\kern-.025em b}\kern-.08emT\kern-.1667em\lower.7ex\hbox{E}\kern-.125emX}}
      
\copyrightyear{2021}
\acmYear{2021}
\setcopyright{acmlicensed}
\usepackage{verbatim}  
\usepackage{algorithm}
\usepackage{algpseudocode}  
\usepackage{subfigure}
\usepackage{multirow}
\usepackage{threeparttable}
\newcommand{\blue}[1]{\textcolor{black}{#1}}
\newcommand{\black}[1]{\textcolor{black}{#1}}
\begin{document}

%
\title{\black{Reducing Redundancy in Data Organization and Arithmetic Calculation for Stencil Computations}}
\renewcommand{\shorttitle}{Reducing Redundancy in Data Organization and\\ Arithmetic Calculation for Stencil Computations}

\author{Kun Li}
\affiliation{%
  \institution{Institute of Computing Technology, Chinese Academy of Sciences} 
  \institution{University of Chinese Academy of Sciences} 
  \city{Beijing}
  \state{China}}
\email{likungw@gmail.com}
\author{Liang Yuan}
\affiliation{%
  \institution{Institute of Computing Technology, Chinese Academy of Sciences} 
  \city{Beijing}
  \state{China}} 
\email{yuanliang@ict.ac.cn} 
\author{Yunquan Zhang} 
\affiliation{%
  \institution{Institute of Computing Technology, Chinese Academy of Sciences} 
  \city{Beijing}
  \state{China}} 
\email{zyq@ict.ac.cn} 
\author{Yue Yue}
\affiliation{%
  \institution{Institute of Computing Technology, Chinese Academy of Sciences} 
  \institution{University of Chinese Academy of Sciences} 
  \city{Beijing}
  \state{China}}
\email{yyue1998@gmail.com}
\author{Hang Cao}
\affiliation{%
  \institution{Institute of Computing Technology, Chinese Academy of Sciences} 
  \institution{University of Chinese Academy of Sciences} 
  \city{Beijing}
  \state{China}}
\email{caohang@ict.ac.cn}
\author{Pengqi Lu}
\affiliation{%
  \institution{Institute of Computing Technology, Chinese Academy of Sciences} 
  \institution{University of Chinese Academy of Sciences} 
  \city{Beijing}
  \state{China}}
\email{lupengqi18s@ict.ac.cn} 

\begin{abstract}
  Stencil computation is one of the most important kernels in various scientific and engineering applications. A variety of work has focused on vectorization techniques, aiming at exploiting the in-core data parallelism. Briefly, they either incur data alignment conflicts or hurt the data locality when integrated with tiling. In this paper, a novel transpose layout is devised to preserve the data locality for tiling in the data space and reduce the data reorganization overhead for vectorization simultaneously. We then propose an approach of temporal computation folding designed to further reduce the redundancy of arithmetic calculations by exploiting the register reuse, alleviating the increased register pressure, and deducing generalization with a linear regression model. Experimental results on the AVX-2 and AVX-512 CPUs show that our approach obtains a competitive performance. 
\end{abstract} 

%
%

%
\keywords{Stencil, Vectorization, Register reuse, Data locality}

\maketitle

\section{Introduction} 

Stencil is one of the most important kernels widely used across a set of scientific and engineering applications. It is extensively involved in various domains from physical simulations to machine learning \cite{sawdey1995design,li2019openkmc,chen2020scene}. Stencil is also included as one of the seven computational motifs presented in the Berkeley View \cite{asanovic2006landscape,asanovic2008parallel,10.1145/3126908.3126920} and arises as a principal class of floating-point kernels in high-performance computing. 

A stencil contains a pre-defined pattern that updates each point in a $d$-dimensional spatial grid iteratively along the time dimension. The value of one point at time $t$ is a weighted sum of itself and its neighboring points at the previous time \cite{10.1145/1989493.1989508,datta2008stencil}. The naive implementation for a $d$-dimensional stencil contains $d+1$ loops where the time dimension is traversed in the outmost loop and all grid points are updated in inner loops. Since stencil is characterized by this regular computational structure, it is inherently a bandwidth-bound kernel with a low arithmetic intensity and poor data reuse \cite{10.1145/3126908.3126920,10.1145/1273442.1250761}.


Performance optimizations of stencils has been exhaustively investigated in the literature. Traditional approaches have mainly focused on either vectorization or tiling schemes, aiming at improving the in-core data parallelism and data locality in cache respectively. These two approaches are often regarded as two orthogonal methods working at different levels. Vectorization seeks to utilize the SIMD facilities in CPU to perform multiple data processing in parallel, while tiling tries to increase the reuse of a small set of data fit in cache. 


Prior work on vectorization of stencil computations primarily falls into two categories.
The first one is based on the associativity of the weighted sums of neighboring points. Specifically, the execution order of one stencil computation can be rearranged to exploit common subexpressions or data reuse at register or cache level 
\cite{Cruz.Araya-polo:toms14,zhao2019exploiting,7161520,rawat2018register,8665800}. 
Consequently, the number of load/store operations can be reduced and the bandwidth usage is alleviated in optimized execution order. 
The second one attempts to deal with the data alignment conflict
\cite{henretty2011data,10.1145/2464996.2467268},
which is the main performance-limiting factor. 
The data alignment conflict is a problem caused by vectorization, \black{where the neighbors for a grid point appears in the same vector registers but at different positions.}
One milestone approach is DLT method (Dimension-Lifting Transpose) \cite{henretty2011data}, \black{and it performs a global matrix transpose to address the data alignment conflict.}


As one of the crucial techniques to exploit the parallelization and data locality for stencils, tiling, also known as blocking, has been widely studied for decades. Since the size of the working sets is generally larger than the cache capacity on a processor \cite{10.1145/3155290}, the spatial tiling algorithms are proposed to explore the data reuse by changing the traversal pattern of grid points in one time step. 
However, such tiling techniques are restricted to the size of the neighbor pattern \cite{10.1145/1273442.1250761,yuan2019tessellating}. Temporal tiling techniques have been developed to allow more in-cache data reuse across the time dimension \cite{10.1145/3126908.3126920}. 


The aforementioned two approaches of stencil computation optimizations
often have no influence on the implementation of each other.
However, the data organization overhead for vectorization 
may degrade the data locality.
Moreover, most of the prior work only focuses on temporal tiling on the cache level.
This only optimizes the data transfer volume between cache and memory, 
and the high bandwidth demands of CPU-cache communication is still
unaddressed or even worse with vectorization. Thus, redundant calculation is performed on the same grid point iteratively due to massive CPU-cache transfers along the time dimension.


In this paper, we first design a novel transpose layout to overcome the input data alignment conflicts of vectorization and preserve the data locality for tiling simultaneously.
The new layout is formed with an improved in-CPU matrix transpose scheme,
which achieves the lower bounds both on the total number of data organization operations and the whole latency.
Compared with conventional methods, the corresponding computation scheme for the new layout requires 
less data organization operations,
whose cost can be further overlapped by arithmetic calculations.

Based on the proposed data layout, a temporal computation folding approach is devised to reduce the redundancy of arithmetic calculations. 
We perform a deep analysis of the expansion for multiple time steps, fold the redundant operations on the same point, and reassign a new weight for it to achieve a multi-step update directly.
An improved in-CPU flops/byte ratio is obtained by reusing registers, and the calculation of intermediate time steps are skipped over to alleviate the increased register pressure. 
Furthermore, we utilize a shifts reusing technique to decrease the redundant computation within the innermost loops, and integrate the proposed approach with a tiling framework to preserve the data locality.
Finally, the temporal computation folding approach is generalized for arbitrary stencil pattern by using a linear regression model.

The proposed scheme is evaluated with AVX-2 and AVX-512 instructions for 1D, 2D, and 3D stencils. The results show that our approach is obviously competitive with the existing highly-optimized work\cite{10.1145/2464996.2467268,henretty2011data,10.1145/3126908.3126920}. 

This paper makes the following contributions:
 


\begin{itemize}

\item We propose an efficient transpose layout
and corresponding vectorization scheme for stencil computation.
The layout transformation utilizes an improved matrix transpose
of the lowest latency.

\item \black{Based upon the new proposed transpose layout, we design a temporal computation folding approach optimized by shifts reusing and tessellate tiling. It aims to reduce the redundancy of arithmetic calculation in time iteration space. }


\item We generalize our approach on various benchmarks, and demonstrate that it could achieve superior performance compared to several highly-optimized work\cite{10.1145/2464996.2467268,henretty2011data,10.1145/3126908.3126920} on multi-core processors.

\end{itemize}

The paper is organized as follows. 
\black{
Section 2 elaborates on the addressed data organization problem and formally describes the proposed vectorization scheme. 
The redundancy elimination of arithmetic calculation in time iteration space and its implementation are discussed in Section 3. }
Section 4 discusses the experimental results. In Section 5, we present the related work and Section 6 concludes the paper.

\section{\label{ttl}Spatial Data Organization}

In this section, we first discuss the drawbacks of existing methods.
Then 
we present a new transpose layout and its corresponding vectorized computation process.
Next, \black{we present an improved transpose implementation in our work.}

\subsection{Motivation}


DLT is a promising method that extremely reduces the data reorganization operations. \black{It performs a global matrix transpose to reconstruct the whole data layout in memory \cite{henretty2011data}. However, the elements in one vector are distant, thus there is no data reuse among them.} Furthermore, DLT suffers from the overhead of explicit transpose operations executed before and after the stencil computation. 
For high-dimensional stencils and low-dimensional in other applications like image processing, the time size is small that makes the global transpose overhead unignorable.
Finally, it's hard to implement DLT transpose in-place and it often chooses to use an additional array to store the transposed data. This increases the space complexity of the code.

Our starting point is the observation of the disadvantages of existing methods. Essentially DLT vectorization format hurts the locality properties as mentioned above. On the contrary, the straightforward multiple load and data reorganization methods load contiguous element in one vector. They lead to the optimal data locality when integrated with a temporal tiling scheme.

These two methods seem to be at two extreme ends of a balance between the number of reorganization operations of data in CPU and the reuse ability of data in cache.
Our scheme seeks to preserve the data locality property
and employs the fundamental idea of DLT to improve the overhead of
data preparation.

\begin{figure}[t] 
  \begin{center}
  \centering
  \includegraphics[width=0.4\textwidth]{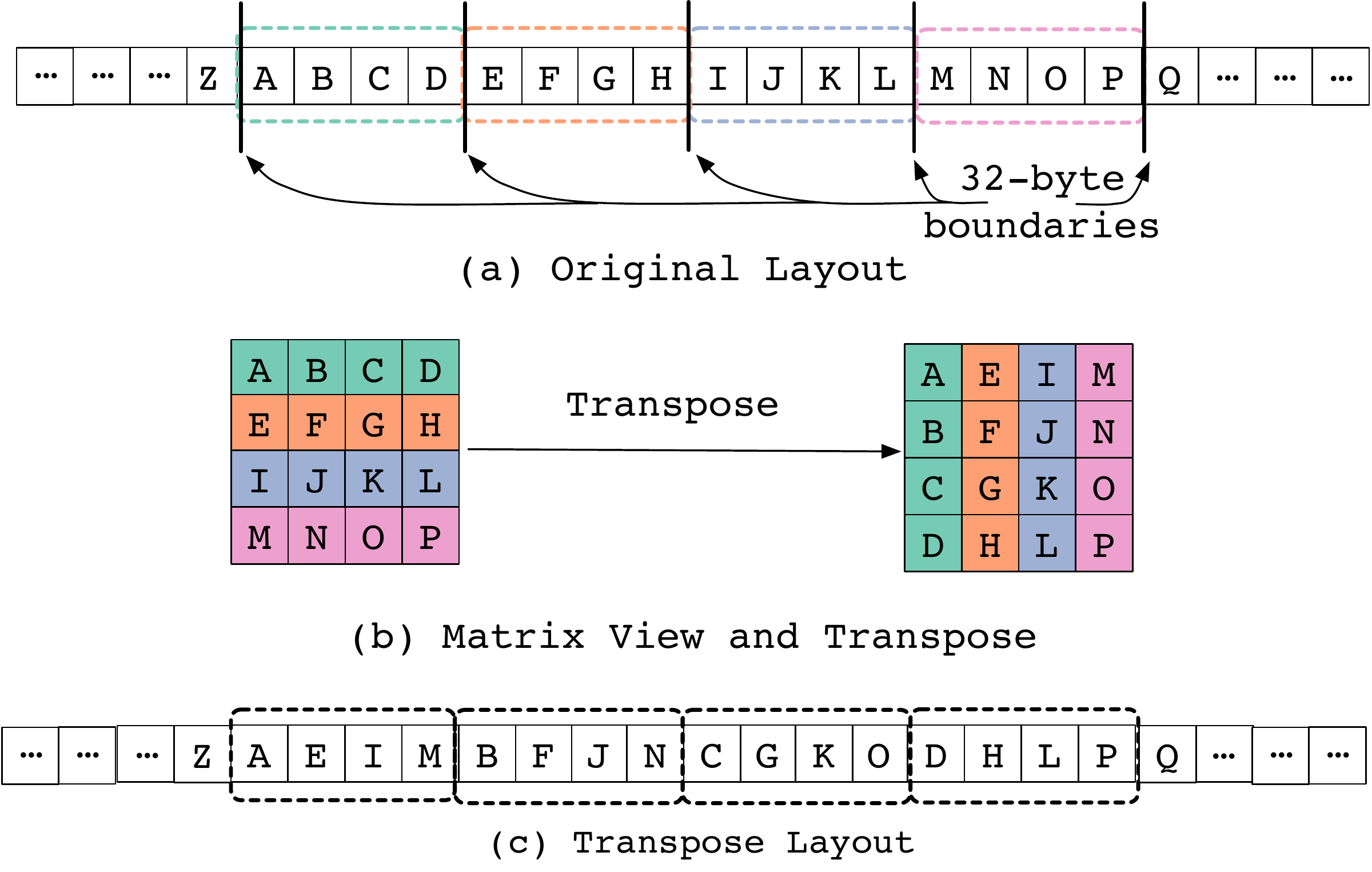}
  \caption{\label{rls}Register Transpose Layout for SIMD vector length of 4.}
  \end{center}
\end{figure}

\begin{figure} 
  \begin{center}
  \centering
  \includegraphics[width=0.3\textwidth]{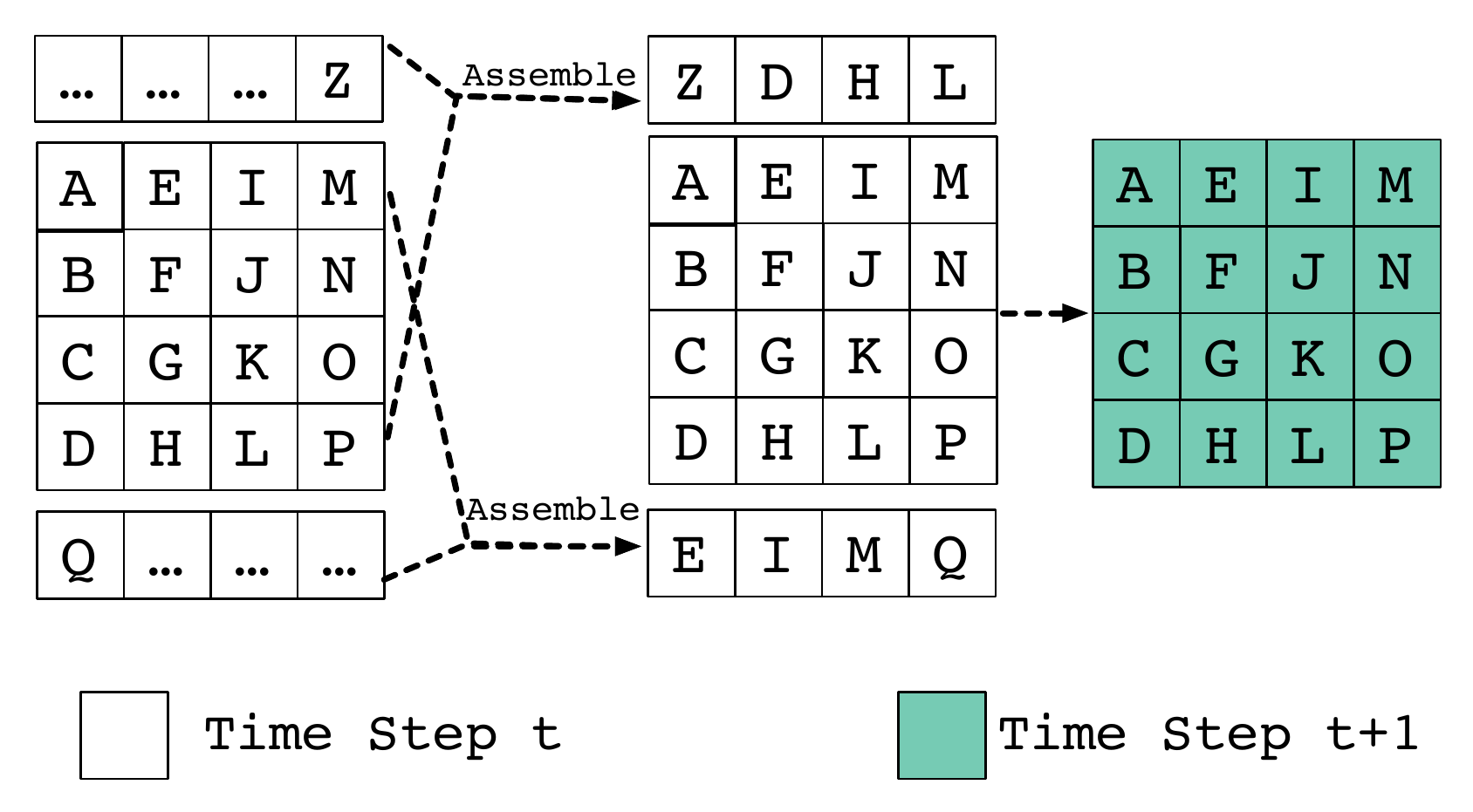}
  \caption{\label{multi} Illustration of stencil computation for transpose layout.}
  \end{center}
\end{figure}

\subsection{\label{subrls}The Transpose Layout}
\paragraph{Locally Transpose}
To preserve the data locality and 
reduce the number of data organization operations,
we apply a matrix transpose to a small sub-sequence of contiguous elements.
Specifically, like the dimension-lifting approach in DLT, 
the one-dimensional view to the sub-sequence is substituted by a two-dimensional
matrix view. To perform vectorization after a matrix transpose,
the column size of the matrix should be equal to $vl$, where $vl$ is the vector length in vector elements.
Figure \ref{rls} illustrates the transpose layout for a one-dimensional stencil with $vl=4$. The matrix transposes of every sub-sequence of $vl*vl$ length is only performed twice before and after the stencil computation respectively.
In the rest of the paper, we also refer to the $vl$ vectors as a \textbf{vector set}.
Note that in the implementation a vector set is always aligned to a 32-Byte boundary.

\paragraph{Vectors Computation}
The stencil computation of the vector set is straightforward
as shown in Figure \ref{multi}. 
The update of one vector set of the 3-point stencil requires two assembled vectors.
One is the left dependent vector of its first vector
and the other is the right dependent vector of its last vector.
Figure \ref{multi} describes the data reorganization of these two vectors.
The first vector is $(A,E,I,M)$ and its left dependent vector is $(Z,D,H,L)$ which is stored in 
two distant vectors in the transpose layout, $(*,*,*,Z)$ and $(D,H,L,*)$.
These two vectors are combined by a blend instruction
followed by a permute operation to shift the components to the right circularly.

We then achieve an efficient vectorization scheme by performing lower-overhead matrix transpose and two data operations per vector set.
There are several considerations for devising this vectorization scheme. 
First, to avoid an additional array that is needed to store the transposed data as in DLT format, it's desirable to complete the matrix transpose in CPU.
The second reason is that transposing a matrix of size $vl*vl$ is cheaper to implement on modern CPU products.
We will present a highly efficient algorithm
for matrix transpose of size $vl*vl$ later.
Moreover, the proposed vectorization scheme avoids data reloads compared with the multiple load method and frequent inter-vector permutations compared with the data reorganization method. The transpose layout could also be applied to higher-order and multidimensional stencils in the same manner.

\subsection{Implementation} 
\begin{figure}
  \label{cases}
  \begin{center}
  \centering  
  \includegraphics[width=0.38\textwidth]{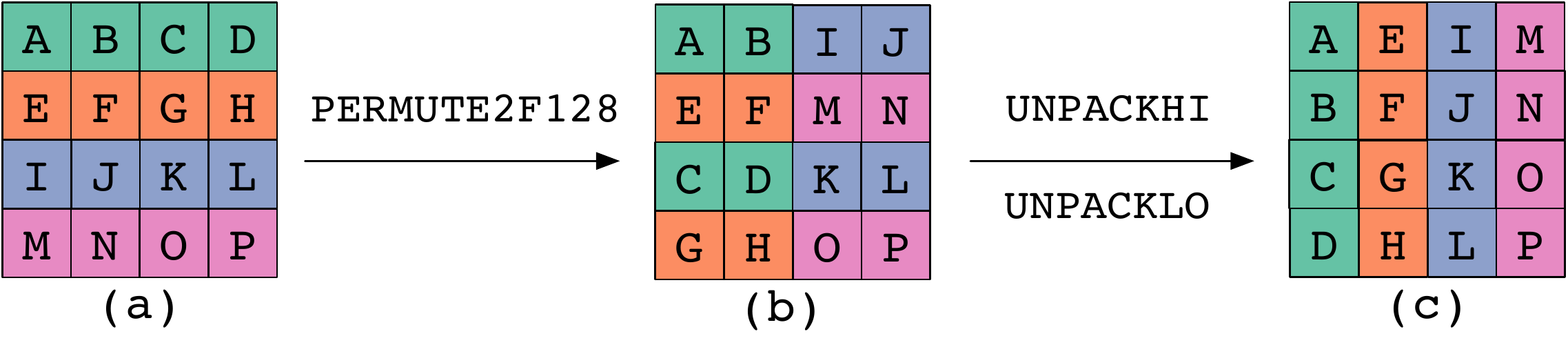}
  \caption{\label{transpose}{Transpose for $double$ type using AVX-2 instructions.} }
  \end{center}
  \vspace{-0.5cm}
\end{figure}

Unlike previous work \cite{henretty2011data} that performs a global dimension-lifted transformation, we only need a transpose on-the-fly for each register set twice throughout the whole process.
Figure \ref{transpose} illustrates the improved version by single-cycle non-parameter U{\scshape npack} instructions. In the first stage, pairs of two vectors with distance $2$, e.g., $(A,B,C,D)$ and $(I,J,K,L)$, exchange data using the P{\scshape ermute2f128} instruction. In the second stage, the pairs of two adjacent vectors, e.g., $(A,B,I,J)$ and $(E,F,M,N)$, swap elements by the U{\scshape npackhi} or U{\scshape npacklo} instruction. In modern CPU architectures, these 8 instructions on 4 vectors can be launched continuously in 8 cycles, which are even less than the sum cost of them (4 cycles for a(n) add/multiply-add instruction). 
Similarly, the transpose by using AVX-512 instructions contains three stages where the last stage consists of in-lane instructions.

\section{Temporal Computation Folding}

\subsection{Overview of Approach}

In general, all grid points are only updated once before the round starts for the next time step in stencil computation. Although most of the existing work utilizes blocking technique \cite{yuan2019tessellating,10.1145/3126908.3126920,bandishti2012tiling} to decrease the data transfers between main memory and cache, there is no in-register data reuse between $m$ successive time loops, where $m$ is called the unrolling factor. 
On the contrary, the straightforward implementation of reusing registers along the time dimension produces massive intermediate results at the time step $t$ to $t+m$, which exacerbates excessive register spilling. 
\begin{figure*}
  \begin{center}
  \centering
  \includegraphics[width=0.95\textwidth]{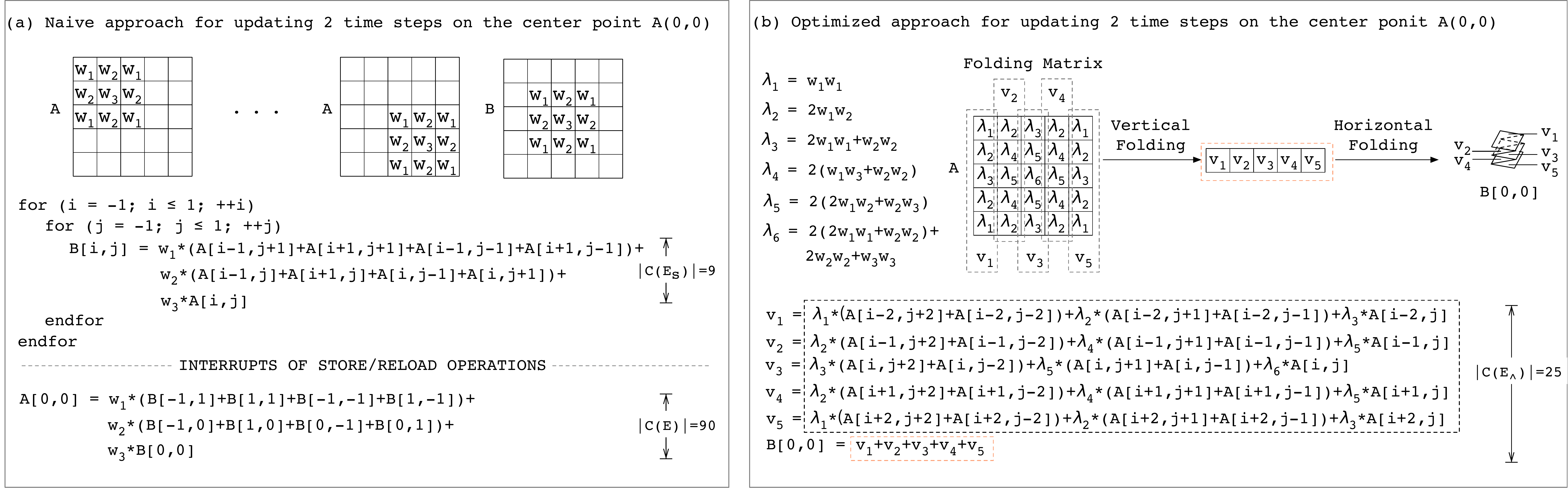}
  \caption{\label{dag}{Illustration of scalar arithmetic expression for the 9-point box stencils with $m=2$. The approach is used to try to minimize the collect $C(E)$ during the computation process.} }
  \end{center} 
\end{figure*} 

The existing work and straightforward implementation represent opposite extremes of reusing register in time iteration space. 
Our approach is to seek a balance that the redundancy of arithmetic calculation is eliminated along the time dimension, and register pressure is alleviated simultaneously. 
To facilitate the process of reducing redundant calculation in time iteration space, we propose a computation folding approach to update the grid points for $m$ time steps directly in registers.  
The computation folding approach is elaborated thoroughly based upon a profitability analysis, and it is optimized with shifts reusing and tessellate tiling to obtain further performance gains. Additionally, we also describe an inductive generalization that promotes our computation folding approach on a wide variety of stencil kernels.

\subsection{\label{scalarpa}Scalar Profitability Analysis}

Figure \ref{dag} shows a scalar arithmetic expression for a representative 9-point box stencil with unrolling factor $m=2$ on the center grid point. A collect $C(E)$ in Equation \ref{ce} is defined to describe the number of instructions (add, multiply, multiply-add, etc.) used for $m$-step updates in the expression $E$.

\begin{eqnarray}
  \label{ce}
  C(E)=\bigcup\limits_{s}\{\langle g, w_g \rangle| \langle g, w_g \rangle \in C(E_s) \}
\end{eqnarray}  
 
For the naive expression in Figure \ref{dag}(a), the center point with neighboring eight grid points are all updated to the state $t+1$ first, and then these updates are swept from registers to memory. When the next iteration for $t+2$ begins, these grid points of $t+1$ are reloaded again. 
To obtain a 2-step updates on the center point, the computing instructions of ten subexpressions are all counted into the collect $C(E)$. In each subexpression $E_s$, a pre-defined weight $w_g$ is assigned on each grid point and then a 9-way addition result is obtained. Since nine distinct point references are engaged for each subexpression, we obtain a $|C(E)|= 10\times|C(E_s)|=90$ for the expression $E$. It is worth noting that redundant arithmetic operations are performed iteratively on the same point in different subexpressions, and store/reload operations incur a costly interrupt during the computation process.
 

For the optimized expression in Figure \ref{dag}(b), weights are all reassigned based on the $m$-step expansion. A new arithmetic expression $\mathbb{E}_{\Lambda}$ is determined by the folding matrix comprised of new weights $\lambda$. The five associative grid points of the same column are folded with $\lambda$ first, and then a horizontal folding is performed to gather the obtained five folded values. Thus, the new collect in Equation \ref{uce} is 25, which is obtained from the computation folding on this point set with each grid point folded by $\lambda_g$.

\begin{eqnarray}
  \label{uce}
  C(\mathbb{E}_{\Lambda})=\{\langle g,\lambda_g \rangle| \text{ grid } g \text{ used in }\mathbb{E}_{\Lambda} \text{ weighted }  \text{ with } \lambda_g \}
\end{eqnarray}

The profitable index is evaluated in Equation \ref{pi}, and a profitable folding means the fraction of the cardinalities on two sets at least exceeds a threshold $\theta \ge 1$. In this case, it gives a net profitable index of $P(E,\mathbb{E}_{\Lambda})=90/25=3.6$ from Equation \ref{pi}. Moreover, the interrupt cost of store/reload operations is also cut entirely in $\mathbb{E}_{\Lambda}$. 

\begin{eqnarray}
  \label{pi}
  P(E,\mathbb{E}_{\Lambda})=\frac{|C(E)|}{|C(\mathbb{E}_{\Lambda})|}\ge \theta
\end{eqnarray}

\subsection{\label{compfold}Vectorized Multi-step Computation}
\begin{figure*}
  \begin{center}
  \centering
  \includegraphics[width=0.9\textwidth]{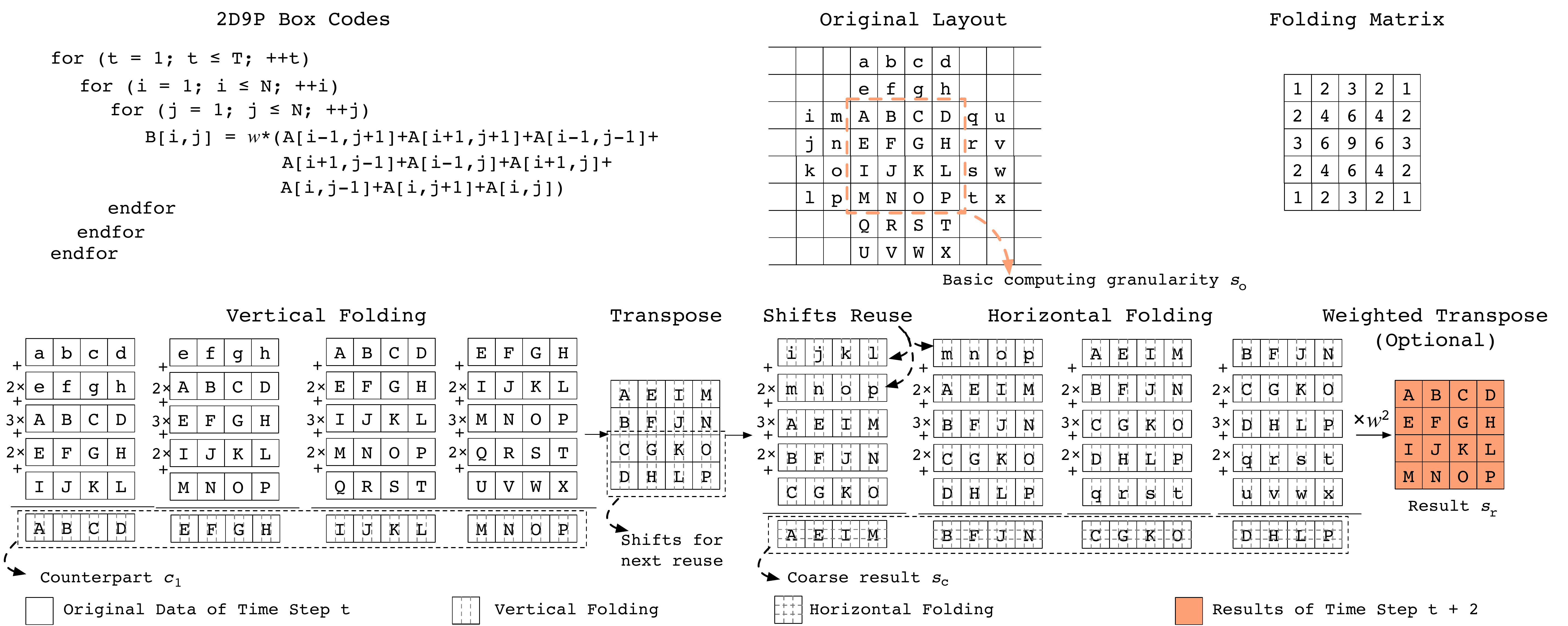}
  \caption{\label{foldings}{Vectorized process of computation folding approach for 9-point box stencils with unrolling factor $m=2$.} }
  \end{center}
\end{figure*} 
 

In this subsection, we take a $2$-step 2D9P box stencil \cite{10.1145/3126908.3126920,yuan2019tessellating} as an example to illustrate the details on our computation folding approach in Figure \ref{foldings}, which is a vectorized process of the optimized arithmetic expression discussed in Section \ref{scalarpa}.

\paragraph{Layout Preparation}
Figure \ref{foldings} depicts the codes and layout preparation for the box stencils. Based on the previous data organization in Section \ref{ttl}, the basic granularity of computation is also constructed as a 4$\times$4 square of grid points denoted as $s_o$. They are loaded from cache to four registers as $v_0$ to $v_3$ respectively at time step $t$. The transpose layout proposed in Section \ref{ttl} is utilized to manipulate the registers, 
and it can be adapted on multidimensional stencils. For example, 1D stencils with length of 4$N$ grid points are viewed as a 4$\times$N 2D grid points, and 3D stencils with volume of $N_x \times N_y \times N_z$ are manipulated as a $N_z$-layer 2D $N_x \times N_y$ slice.

\paragraph{Vertical Folding}
Vertical Folding is performed first to collect neighbor points in the same column. A new square of grid points derived from $s_o$ with vertical folding is called as a counterpart. Typically a $m$-step update contains $m+1$ counterparts at most. Equation \ref{vfold} describes how each counterpart is obtained by vertical folding:  
\begin{eqnarray}
  \label{vfold}
  {v^{(n)}_i}={\sum}_{t=-m}^{m}\lambda_t^{(n)} \cdot {v_{i+t}}
\end{eqnarray}
, where $\lambda$ is the reassigned weight; superscripted $n$ is the counterpart number. For example, the weights for the first counterpart are reassigned as $\boldsymbol{\lambda^{(1)}}$=\{1, 2, 3, 2, 1\} by the folding matrix shown in Figure \ref{foldings}. According to Equation \ref{vfold}, each $v_i^{(1)}$ in the first counterpart $c_1$ is calculated by performing a sum on $v_{i-2}$, 2$v_{i-1}$, 3$v_{i}$, 2$v_{i+1}$, and $v_{i+2}$.

\paragraph{Horizontal Folding}
With vertical folding completed, a local transpose is performed subsequently for further horizontal folding to collect the folded values in the same row:
\begin{eqnarray}
  \label{horifold}
  {v_i^,}={\sum}_{t=-m}^{m} {{v^{(c-|t|)}_{i+t}}}
\end{eqnarray}
, where $c$ is the total number of counterparts. Since reassigned weights for the other two counterparts $c_2$ and $c_3$ are represented by $\boldsymbol{\lambda^{(2)}}=2\boldsymbol{\lambda^{(1)}}$=\{2, 4, 6, 4, 2\} and $\boldsymbol{\lambda^{(3)}}=3\boldsymbol{\lambda^{(1)}}$=\{3, 6, 9, 6, 3\}, the Equation \ref{horifold} could be expanded as:
\begin{eqnarray}
  \label{h_fold_equation}
  \begin{aligned}
  {v_i^,}&={{v^{(1)}_{i-2}}}^{}+{{v^{(2)}_{i-1}}}^{}+{{v^{(3)}_{i}}}^{}+{{v^{(2)}_{i+1}}}^{}+{{v^{(1)}_{i+2}}}^{}\\
  &={{v^{(1)}_{i-2}}}^{}+2{{v^{(1)}_{i-1}}}^{}+3{{v^{(1)}_{i}}}^{}+2{{v^{(1)}_{i+1}}}^{}+{{v^{(1)}_{i+2}}}^{}\\
  \end{aligned}
\end{eqnarray}.
Thus, a coarse result $s_c$ for 2-step updates on a point square $s_o$ is obtained by only utilizing the square $c_1$.

\paragraph{Weighted Transpose}

Horizontal folding is followed by a weighted transpose at last. Conventionally the stencil of Jacobi style is implemented with two arrays \cite{bondhugula2008practical,10.1145/3126908.3126920}, storing the value at odd and even time respectively. Therefore, the local transpose is optional here, and the result $s_r$ could be organized to the original layout by the transpose in horizontal folding alternately. The whole vectorized process is performed by using computation folding only on $c_1$ in practice, which correlates with the $v_1$ of scalar expression in Figure \ref{dag}. Thus the $|C(\mathbb{E}_{\Lambda})|$ in Equation \ref{pi} is further decreased to 9, and we obtain a profitable index $P(E,\mathbb{E}_{\Lambda})=10$ theoretically.

\subsection{Optimization}
In this subsection, we present additional optimizations that we utilized in our approach. 

\paragraph{Shifts Reusing}
\begin{figure}
  \begin{center}
  \centering
  \includegraphics[width=0.3\textwidth]{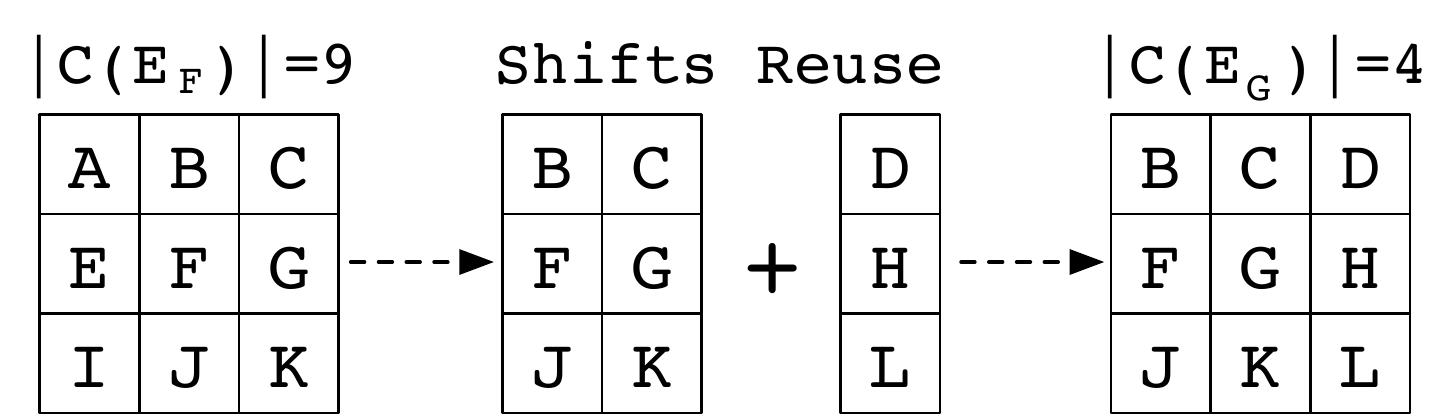}
  \caption{\label{shifts}{Illustration of shifts reusing in stencil computation.} }
  \end{center} 
\end{figure} 

Figure \ref{shifts} depicts a brief sketch of the scalar 1-step stencil computations between two adjacent grid points F and G in data space. It can be recognized from Figure \ref{shifts} that there is potential for reusing shifts within the successive stencil computation from grid point F to G, and this gives us another reuse profitability of 2.25 by Equation \ref{pi}. For our approach in Figure \ref{foldings}, the last two vectors of transposed counterpart $c_1$ in each iteration can be reused as shifts between computing squares. Therefore, the optimization of reducing reloads across squares is enabled by utilizing the same data collected in the last round as input to be computed together, which contributes to further performance gains. 

\paragraph{Tessellate Tiling}
  
\begin{figure}[t]
  \label{cases}
  \begin{center}
  \centering
  \includegraphics[width=0.45\textwidth]{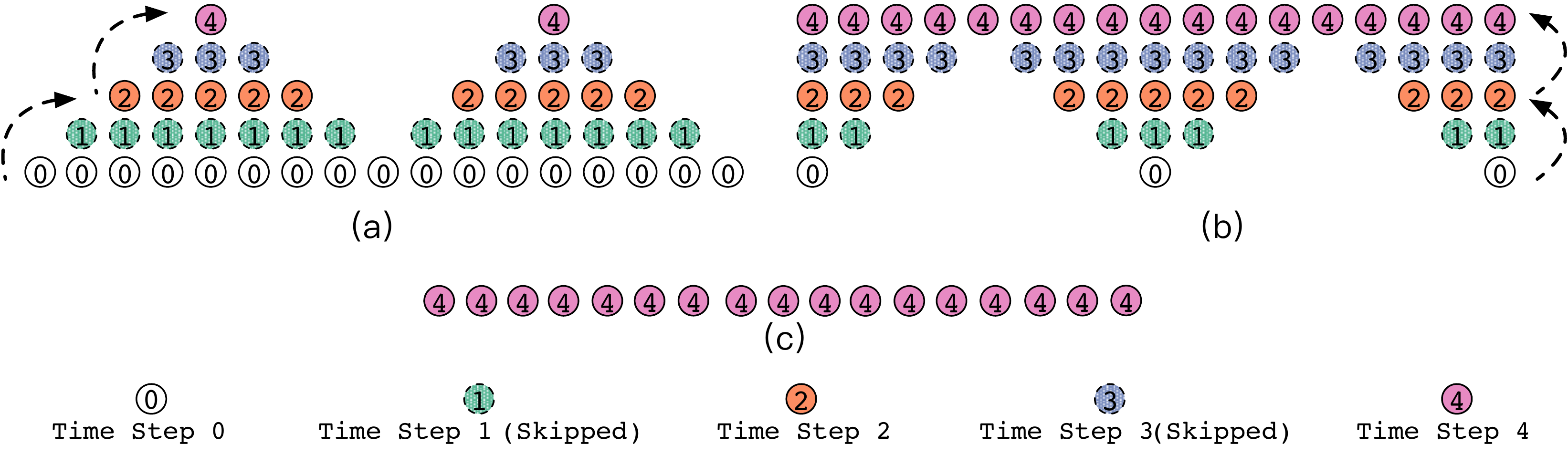}
  \caption{\label{tiles}Our scheme integrated with tessellate tiling for 3-point stencils of unrolling factor $m$=2.}
  \end{center}
\end{figure}

We also present a combination of our scheme and tessellate tiling \cite{yuan2019tessellating} in our work.
The tessellate tiling can be viewed as a tessellation in iteration space by utilizing shaped tiles. Figure \ref{tiles} (a) and Figure \ref{tiles} (b) illustrate the our scheme integrated with tiling for a 3-point stencil. The iteration space is tessellated by triangles and inverted triangles in alternative stages. Thus, concurrent execution is processed by two stages which are started in each triangle with a given time range first, followed closely by the execution of inverted triangles over the same time range concurrently. For the example in Figure \ref{tiles} (a), the new state of each triangle contains (0,1,2,3,4,3,2,1,0) where the center element is updated four steps and its neighbors are updated fewer steps proportional to the distance with the center element. To make all elements updated with the same steps, two half parts from adjacent triangles constitute new inverted triangles and the elements are updated with the state (4,3,2,1,0,1,2,3,4). As Figure \ref{tiles} (c) shows, all elements are updated to four steps by adding the projection of the triangles with inverted triangles. Moreover, the odd time steps are skipped over when computation folding approach is used with $m$=2.
The tessellate tiling could also be integrated into multidimensional stencil computations. For a $d$-dimensional stencil, tessellation in iteration space contains $d+1$ stages. The spatial space in stage $i$ is tessellated by $tiles_i$ ($0\le i\le d$). $tiles_1$ is a hypercube (typically a line segment in 1D, square in 2D, cube in 3D). $tiles_{i+1}$ is built by recombining the sub-tiles split from adjacent $tiles_i$ along some dimensions. Therefore, concurrent execution for different tiles is enabled over a given time range without redundant computation by the integrated tiling. The register transpose layout and time loop fusion also make it feasible to achieve multiple time steps computation in registers over the tiles efficiently without reloading operations.

\subsection{Generalization}
Although a standard 2D9P box stencil is illustrated as an example in Section \ref{compfold}, the computation folding approach can be extended in arbitrary stencil pattern by parameter tuning.
It is observed that the computing expressions of different stencils reflect in the distinct $\boldsymbol{\lambda}$ in folding matrix, and various counterparts are obtained by it. However, the $\boldsymbol{\lambda^{(m)}}$ for counterpart $c_m$ multiple of the other $\boldsymbol{\lambda^{(n)}}$ for counterpart $c_n$ is not guaranteed in some cases. Thus, we drill down into the computing process of these counterparts, and propose a linear regression model \cite{cosenza2017autotuning,chen2020scene} to minimize $|C(\mathbb{E}_{\Lambda})|$ by reusing the available counterparts. For the $n^{th}$ counterpart ${c_n}$, it can be obtained by Equation \ref{linear}: 
\begin{eqnarray}
  \label{linear}
  \begin{aligned}
  c_n&=\omega_{n-1}^{} c_{n-1}+\omega_{n-2}^{} c_{n-2}+...+\omega_{2}^{} c_2+\omega_{1}^{} c_1+b_n\\
  &=\boldsymbol{\omega^{}}\boldsymbol{c}+b_n
  \end{aligned}
\end{eqnarray}
, where $\boldsymbol{c}$ is the vector of counterparts; $\boldsymbol{\omega^{}}$ is the weight parameters for the counterpart $c_n$; $b_n$ is a bias for fine tuning results by the original square $s_o$. For convenience, we define a function in Equation \ref{maps} that maps the solving cost of counterparts to collects:
\begin{eqnarray}
  \label{maps}
  |C(\mathbb{E}_{\Lambda})|=\phi(\boldsymbol{c})
\end{eqnarray}, 
and the goal is to find the model parameter $\boldsymbol{\omega}$ which minimizes $|C(\mathbb{E}_{\Lambda})|$ most. The search problem is then converted into a minimization problem on the opposite of the difference between optimized $\phi(\boldsymbol{c^*})$ and original $\phi(\boldsymbol{\hat{c}})$, and the cost function $J$ is formally described in Equation \ref{cost}:
\begin{eqnarray}
  \label{cost}
  \begin{aligned}
  &J(\boldsymbol{\omega_i},b_i)=\sum_{i=1}^{n}(\phi(c_i^*(\boldsymbol{\omega_i},b_i))-\phi(\hat{c_i}))^2
\end{aligned}
\end{eqnarray}.  
The parameters of $\boldsymbol{\omega_i}$ and $b_i$ are subject to a constraint that a correct result is produced by them.
Note that ${\hat{c_i}}$ is computed by using $s_o$ only without any counterpart reuse. For example, the counterpart $c_2$ and $c_3$ in Section \ref{compfold} are both required to be recomputed by Equation \ref{vfold} with $\boldsymbol{\lambda^{(2)}}$ and $\boldsymbol{\lambda^{(3)}}$ to obtain ${\hat{c_2}}$ and ${\hat{c_3}}$. Each obtained ${\hat{c_i}}$ is accumulated to the results and then evicted to alleviate the register pressure. With $c_i^*$ optimized by machine learning algorithm, we obtain the weight vector $\boldsymbol{\omega_2}$=(2) for $c_2$ and $\boldsymbol{\omega_3}$=(0, 3) for $c_3$ in this case respectively. An asymmetric box stencil of 9 different weights is evaluated in Section \ref{eval}, demonstrating the benefits of our approach driven by machine learning algorithm.

\section{\label{eval}Evaluation}

In this section, we evaluate our proposed scheme on varied stencils used for real-world applications with AVX-2 and AVX-512 instructions. 

\subsection{Setup}

\begin{table}[b]
  \caption{Parameter description for stencils used in experiments}\label{parameters}
  \renewcommand\tabcolsep{4.0pt} 
  \centering\begin{tabular}{cccc}
  \hline
  Type & Pts & Problem Size & Blocking Size\\ \hline
  1D-Heat&3    & 10240000$\times$1000 & 2000$\times$1000  \\
  1D5P&5    & 10240000$\times$1000 & 2000$\times$500   \\
  APOP&6    & 10240000$\times$1000 & 2000$\times$500   \\
  2D-Heat&5    & 5000$\times$5000$\times$1000 &  200$\times$200$\times$50                \\
  2D9P&9    & 5000$\times$5000$\times$1000 & 120$\times$128$\times$60                \\
  Game of Life&8    & 5000$\times$5000$\times$1000 & 200$\times$200$\times$50                \\
  GB&9    & 5000$\times$5000$\times$1000 & 200$\times$200$\times$50                \\
  3D-Heat&7    &  400$\times$400$\times$400$\times$1000   &20$\times$20$\times$10                     \\
  3D27P&27   &   400$\times$400$\times$400$\times$1000           &  20$\times$20$\times$10                    \\ \hline
  \end{tabular}
\end{table} 

Our experiments were performed on a machine composed of two Intel Xeon Gold 6140 processors with 2.30 GHz clock speed (turbo boost frequency of up to 3.70 GHz), which owns 36 physical cores organized into two sockets. Each core contains a 32KB private L1 data cache, a  1 MB private L2 cache, and a unified 24.75MB L3 cache. AVX-512 instruction set extension is supported and it's able to conduct operations for 8 double-precision floating-point data in a SIMD manner, which yields a theoretical peak performance of 73.6 GFlop/s/core (2649.6 GFlop/s in aggregate).

Since the recent tiling technique (denoted as tessellation) proposed by Yuan \cite{10.1145/3126908.3126920} and the nested/hybrid tiling technique (denoted as SDSL, which is the name of the software package.) presented by Henretty \cite{10.1145/2464996.2467268} outperform the other stencil research like Pluto \cite{bondhugula2008practical,bandishti2012tiling} and Pochoir \cite{10.1145/1989493.1989508}, we take them as two bases of our work, which are vectorized by the compiler and DLT methods, respectively.
All programs were compiled using the ICC compiler version 19.0.3, with the '-O3 -xHost -qopenmp -ipo' optimization flags.

The detailed parameters for stencils used in experiments are described in Table \ref{parameters}, which consists of three star stencils (1D-Heat, 2D-Heat, and 3D-Heat) and three box stencils (1D5P, 2D9P, and 3D27P) corresponding to the references \cite{10.1145/3126908.3126920,10.1145/2464996.2467268}. Star and box equations are symmetric examples that can represent a wide variety of stencil kernels. Moreover, we also collect a series of classic benchmarks used in real-world applications \cite{lecture_stencil,bandishti2012tiling,10.1145/3126908.3126920}:
\begin{itemize}
  \item APOP is a 1D3P stencil from two different input arrays to calculate the American put stock option pricing.
  \item The Game of Life is a cellular automaton proposed by Conway, and the update of each grid depends on all 8 of its neighbors.
  \item GB (general box) is an asymmetric 2D9P stencil that contains 9 different weights on the grid points in the computation process.
\end{itemize}  
   
The default value of total time steps is 1000 or 200 in the references. Thus, we fix it as a larger value of 1000 in our experiments.
Other parameters of each stencil are also fine-tuned based on references work to guarantee that the peak performance for all methods could be reached exactly. 
Since the performance is sensitive to the stencil parameters, significant efforts are required in automatic tuning and this will be done separately as future work.

\subsection{\label{single_thread}Sequential Block-free Results} 

\begin{figure}
  \begin{center}
  \centering
  \includegraphics[width=0.5\textwidth]{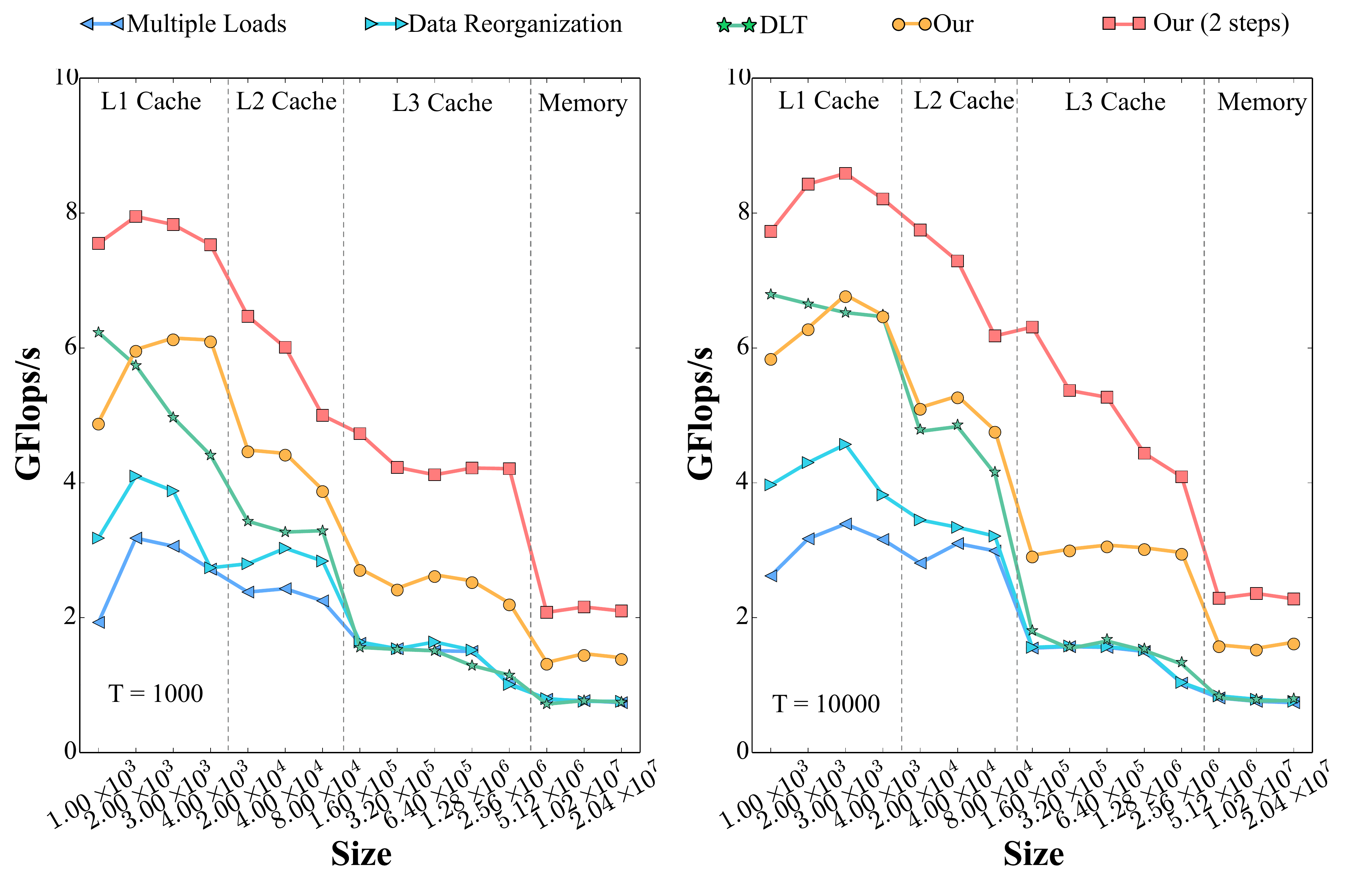}
  \caption{\label{bench}Absolute performance comparison for tested methods in single-thread blocking-free experiments. The results are shown separately with different total time steps.}
  \end{center} 
\end{figure}

In this subsection, we present performance results of varied methods across problem sizes ranging from L1 cache to main memory with a single thread. 
The spatial and temporal blocking method is not applied to them for
examining the pure improvements on various storage levels. 
The multiple loads
and data reorganization methods represent a class of auto-vectorization in modern compilers \cite{10.1145/3126908.3126920}. DLT is the dimension-lifting transpose approach designed by Henretty \cite{henretty2011data}.
All the methods are implemented by hand-written codes optimized with the appropriate strategies such as alignment and loop unrolling to ensure fairness.

Figure \ref{bench} shows the performance comparison of our methods with the other three methods. The results are illustrated separately in two subfigures on the basis of the total time steps $T$. It can be seen that our method updating two time steps outperforms others apparently in both experiments, which demonstrates the effectiveness of the improvement of the flop/byte ratio.
Our method without time loop fusion also achieves better performance results than the hand-written DLT in most cases. The performance has a decrease at the size of 1000 in L1 cache. This can be attributed to the cheaper dimension-lifting transpose operation in small size for DLT. 
The multiple loads method exhibits the worst performance among them due to the overhead caused by redundant loads. Moreover, the performance drops apparently as the problem size moves from L1 cache to the memory hierarchy, which is mainly caused by the cost of data transfers.
%

\begin{table}[htbp]
  \centering\small
  \renewcommand\tabcolsep{3.5pt} 
  \begin{threeparttable}
  \caption{\label{tab:bench}Performance improvements on different storage level in single-thread blocking-free experiments}
  \begin{tabular}{lcccccc}
  \toprule &Multiple&Data  & \multirow{2}{*}{DLT} & \multirow{2}{*}{Our} &\multirow{2}{*}{Our (2 steps)}\\ 
  &Loads&Reorganization&&&\\\midrule
  \multirow{1}{*}{L1 Cache} &1.00x& 1.28x  & 2.06x & 2.16x & 2.83x\\
  \multirow{1}{*}{L2 Cache} &1.00x& 1.11x  & 1.37x & 1.80x& 2.46x \\
  \multirow{1}{*}{L3 Cache} &1.00x& 1.01x& 0.95x& 1.73x& 2.95x \\
  \multirow{1}{*}{Memory} &1.00x& 1.00x &1.01x & 1.81x & 2.68x \\\midrule
  \multirow{1}{*}{Mean}&1.00x&1.11x&1.35x&1.98x&2.79x\\
  \bottomrule \end{tabular} \small
  \end{threeparttable}
\end{table}

To further investigate the effect of total time steps $T$, we perform a tenfold increase on the default value to $T=10000$, which is illustrated in Figure \ref{bench} (b). It can be observed that the performance trends of $T=10000$ are still largely consistent with the results in Figure \ref{bench} (a). However, the performance of our method falls slightly behind DLT in L1 cache, and this performance anomaly is primarily due to the diluted dimension-lifting transpose cost by overly long time steps. Notably, only the performance of DLT in L1 cache drops gradually as problem size increases for both results in Figure \ref{bench}, which is resulted from a costly data layout transformation and indicates a potential bottleneck for cache-blocking. 

We report the detailed results on the relative improvements of absolute performance for the time steps of 1000 on different storage levels in Table.\ref{tab:bench}. The performance improvement of the series of our methods is the largest one in each case, which is unconstrained to the storage level. This reflects the best performance again and corresponds to the results of Figure \ref{bench}.

\subsection{\label{mce}Multicore Cache-blocking Experiments}
\begin{figure*}
  \begin{center}
  \centering
  \includegraphics[width=0.87\textwidth]{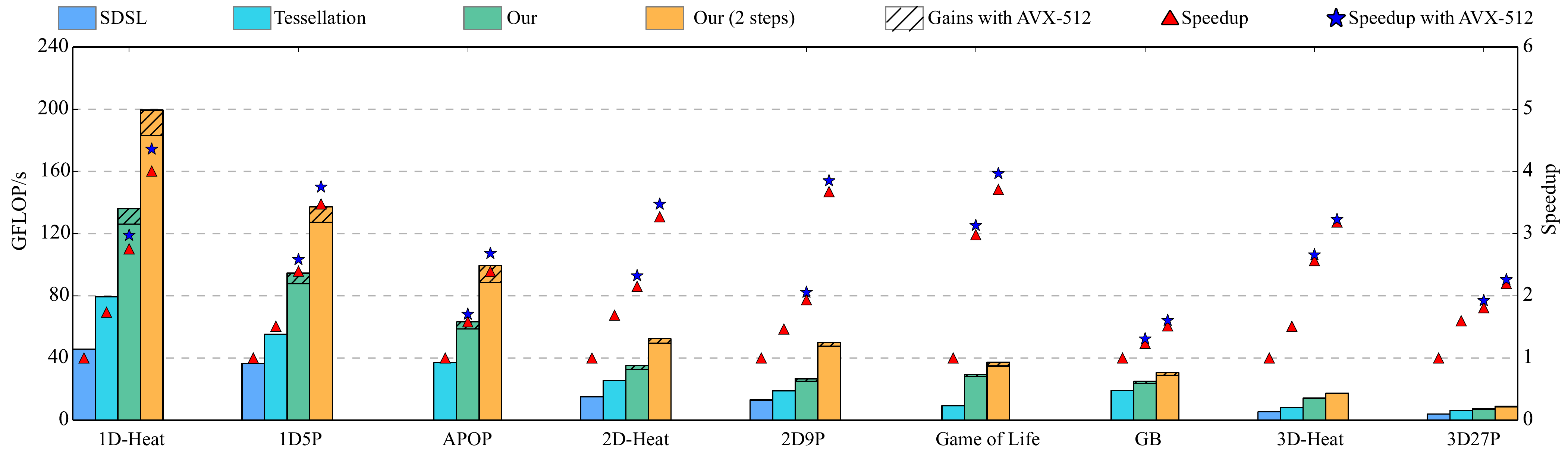}
  \caption{\label{parablock}Performance comparison and speedups for varied methods in multicore cache-blocking experiments.}
  \end{center}
\end{figure*} 

In this subsection, we present the experiments that exhibit the benefits of our methods with the temporal blocking and parallelization scheme. The SDSL employs a split tiling technique (nested tiling in 1D, hybrid tiling for higher dimensions) to achieve temporal blocking \cite{henretty2011data}. The tessellate tiling technique utilized auto-vectorizing supported by the compiler \cite{10.1145/3126908.3126920}.

Figure \ref{parablock} shows the absolute performance comparison and speedups of the different benchmarks optimized by the blocking techniques. 
Since some benchmarks are not supported by SDSL, the speedups of each group are relative to the base which is annotated with the speedup value of 1. 
Taking all stencils with AVX-2 instructions into account, remarkable performance improvements are observed from our method updating two time steps, demonstrating that our vectorization scheme provides a significant benefit in a large problem size compared to the referenced work. 
Moreover, the optimization with AVX-512 instructions could obtain further performance gains. The performance of SDSL is inferior to tessellation, which is resulted from the blocking technique constrained to its data layout.

A closer look at Figure \ref{parablock} indicates that the performance is relative to the shape, dimension, and weight of the stencils. For star-shaped stencils, higher performance improvements are obtained compared to the box-shaped due to fewer neighbor points. For lower-dimensional stencils, much higher reuse is achieved on the loaded inputs, which exhibits better performance. For the real-world stencils, we observed that the performance improvement obtained in GB benchmark is not prominent. 
This is mainly caused by the different weights for each input point in this asymmetric pattern. Although symmetric stencils are preferred in most real-world stencils, the GB benchmark can be viewed as a stress testing on our methods.

\subsection{\label{sca}Scalability}

We also evaluate the scalabilities of our schemes and benchmarks. The detailed parameters are given in Table \ref{parameters}, where all problem sizes exceed the L3 cache. 
Since our tiling framework is the same as the tessellation scheme, 
the performance improvements of our method with respect to the tessellation method are fully derived from the vectorization.
 
\begin{figure*}
  \begin{center}
  \centering
  \includegraphics[width=0.84\textwidth]{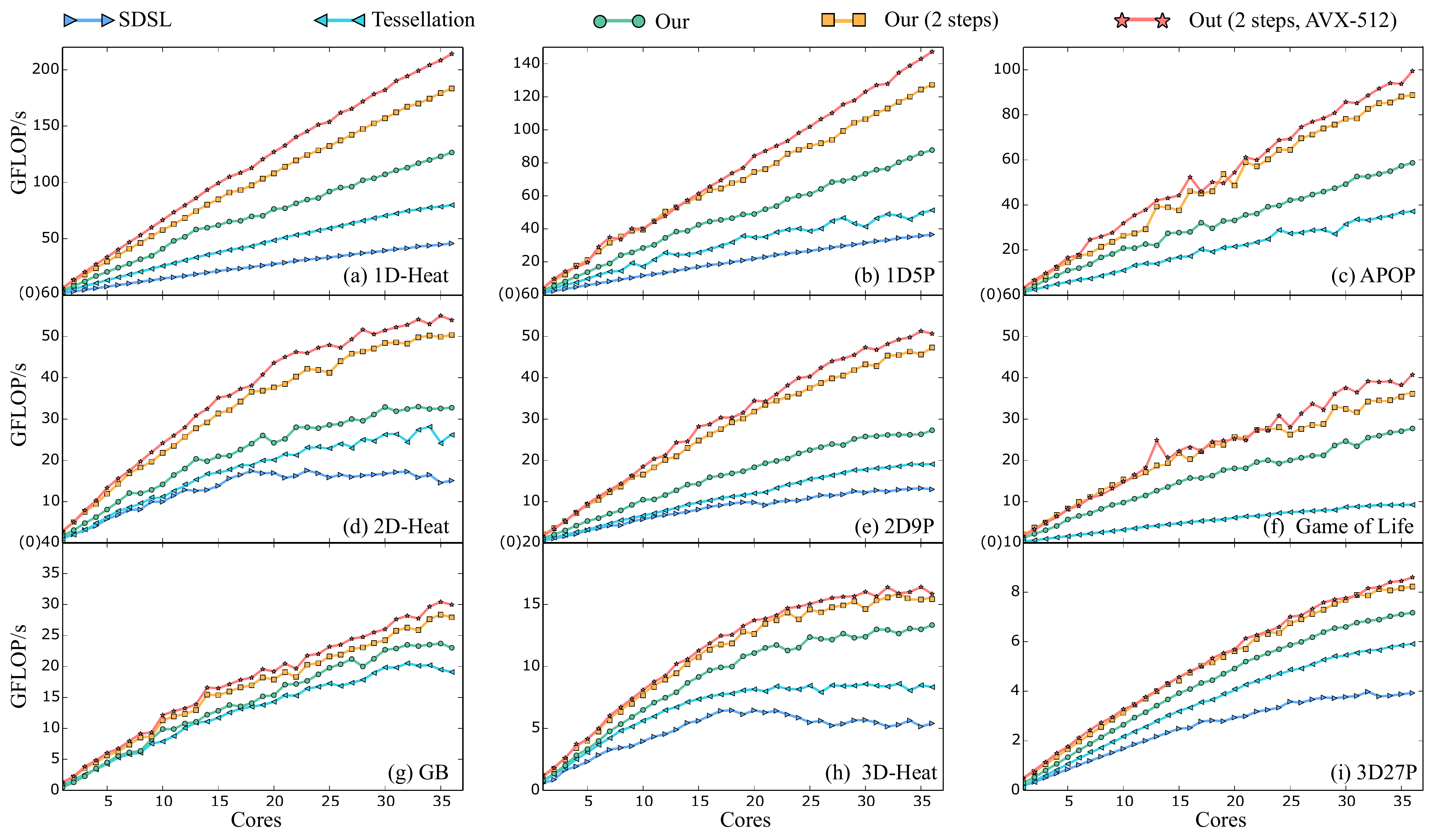}
  \caption{\label{scalings}\blue{\black{Scalability} for stencils of various orders with different dimensions in a multicore environment.}}
  \end{center}
\end{figure*}

It can be observed in Figure \ref{scalings} that our method could obtain the highest performance while the SDSL performs the lowest performance. In 1D3P stencils, all these methods achieve nearly linear scaling on both instruction sets and the proposed temporal computation folding provides a significant improvement. With the increase of the problem dimension, the scalability for all methods drops as a result of the inherent complexity for multidimensional stencil computations. Similarly, the overall performance of high-order stencils also falls behind the corresponding low-order results, which is resulted from complex data access patterns in high-order stencils. Compared to the results implemented with AVX-2 instructions, the performance of AVX-512 optimization shows a further increase.

\begin{table*}[t]
  \centering\small
  \renewcommand\tabcolsep{6pt} 
  \caption{\label{tab:pir}Speedup over single core for different stencils}
  \begin{tabular}{ccccccccccc}
  \toprule &{Method}& {1D-Heat}  & {1D5P} & {APOP} &{2D-Heat}& {2D9P} & {GB} &{Game of Life}  & {3D-Heat} & {3D27P} \\ \midrule
  
  \multirow{5}{*}{Speedup} &SDSL \cite{henretty2011data}& 29.2x & 27.9x& -&18.5x&22.1x&- &-&15.3x &18.7x \\
  &Tessellation \cite{10.1145/3126908.3126920}& 29.8x & 25.4x  &29.5x    & 25.0x   & 29.1x &22.0x  & 23.2x &17.5x&19.2x \\ 
  &Our& 30.8x &30.4x  &30.1x& 30.7x & 30.4x  &23.9x &24.2x &22.1x&24.7x\\
  &Our (2 steps)& \blue{32.1x}&\blue{30.9x}&\blue{31.2x}&\blue{29.1x}&\blue{29.3x}&\blue{22.8x}&\blue{ 24.7x}&\blue{ 22.7x}&\blue{24.9x}\\
  &Our (2 steps, AVX-512)& \blue{31.6x}&\blue{28.7x}&\blue{29.2x}&\blue{28.7x}&\blue{29.6x}&\blue{23.2x}&\blue{24.5x}&\blue{ 21.9x}&\blue{24.9x}\\
  \bottomrule \end{tabular} \small
\end{table*}

The speedup for each method with 36 cores is given in Table \ref{tab:pir}.
For scalability, our method obtains a 24.9x speedup while the value of SDSL is only 18.7x for 3D27P, which indicates a sustainable performance for our method in multidimensional stencils. Additionally, the largest speedup in each stencil column again corresponds to the performance shown in Figure \ref{scalings}, where our method outperforms others in most cases. Thus, our method could obtain a substantial performance improvement in all experiments.

\subsection{Discussion} 

In this subsection, we provide an analysis of the performance on various configurations in previous experiments to tease out the contributions from different aspects of our proposed scheme.

Sequential block-free experiments examine a variety of vectorization methods and demonstrate that our scheme with multiple time steps updating can achieve a considerable 3.32x improvement on average compared with the multiple loads method. Subsequently, the performance gains for larger time steps are still significant and consistent with the results of the small time steps.
Moreover, DLT method is more appropriate only on the relatively small size and long time steps, and this is partly explained by the performance penalty associated with additional dimension-lifting transpose in memory. Since the problem size ranges from L1 cache to main memory, clear insights are provided that the overall performance trends drop consistently with the various memory hierarchy.

Multicore cache-blocking experiments conduct stencil cases with 36 cores, and an average 2.37x speedup is obtained by our method on the basis of SDSL. Due to the reduced arithmetic calculations by our time loop fusion, our method updated with two time steps is proved effective and achieves a further 3.29x speedup. 
Typically many modern processors contain AVX-512 instructions capable of performing on larger 512-bit registers, thus we also study the influence of them. The results reveal that they could contribute to better performance for our methods especially on 1D and 2D stencils. It is worth noting that the frequency reduction called throttling happens on CPUs when heavy AVX-512 instructions are involved. The slowdown is even worse with more cores employed. For example, the turbo frequency of the experimental machine drops significantly from 3.70 GHz to 3.00 GHz when active cores are expanded to full 18 on each processor \cite{xeon_gold}. AVX-512 implementation has a further decrease to 2.10 GHz, which can be blamed for the mediocre performance in 3D stencils. The overall trends are in accord with the sequential block-free experiments, and our method updated with two time steps outperforms others obviously.

The scalability experiments demonstrate that our vectorized scheme leveraging tessellate tiling successfully outperforms the referenced fastest multicore stencil work to date across a broad variety of configurations. Constrained to its specific data layout, DLT is slower than other methods. Since multidimensional or high-order stencils are more compute-intensive, more dependency data are loaded into cache while they are not fully utilized to perform their own stencil computation. Thus, the overall performance for each method falls gradually with the increasing dimensions or orders, and our method could still obtain a better performance.

\section{Related Work}      
 
Research on optimizing stencil computation has been intensively studied \cite{datta2008stencil,kamil2006implicit,meng2009performance,venkatasubramanian2009tuned,strzodka2010cache}, and it can be broadly classified as optimization methods to boost the computation performance, enhance the data reuse, and improve the data locality.

Vectorization by using SIMD instructions is an effective way to improve computation performance for stencils. Henretty proposes a new method DLT \cite{henretty2011data,10.1145/2464996.2467268} to overcome input data alignment conflicts at the expense of a dimension-lifting transpose, which makes it infeasible to perfectly utilize the tiling technique as a result of its spatially separated data elements \cite{10.1145/1273442.1250761}. Essentially DLT can be viewed as the combination of strip-mining (1-dimensional tiling) and out-loop vectorization \cite{10.1145/2464996.2467268}. Specifically, the original innermost loop traverses the corresponding dimension from $1$ to $N$. In DLT the loop is transformed to a depth-2 loop nest where the size of the outer loop equals the vector length $vl$ and the inner loop processes each subsequence of length $N/vl$. Note that the strip-mining was also introduced for vectorization \cite{allen2002optimizing}. However, the conventional usage is to make the size of the innermost loop be the vector length and substitute it by a vector code. Furthermore, the in-place matrix transpose used for vectorization in our work has been widely studied and a kernel of 4$\times$4 matrix transpose consists of two stages basically. Hormati splits the vector register to some 128-bit lanes \cite{hormati2010macross}, and the lane-crossing instructions for $double$ incur a longer latency, typically 3 to 4 cycles. Zekri \cite{zekri2014enhancing} use the in-lane instructions in four stages only for $float$ type. Springer\cite{springer2017ttc} utilize S{\scshape huffle} and P{\scshape ermute2f128} instructions for $double$ type in two stages, while it requires 8 integers as instruction parameters.

Data reuse has also been extensively recognized and exploited. Prior work \cite{zhao2019exploiting,rawat2018register,8665800,stock2014framework} on optimizing the order of execution instructions could decrease loads/stores operations to relieve the register pressure, while only the individual element in each vector could be reused. Basu designs a vector code generation scheme to reuse several vectors in the computation process, and it is constrained to constant-coefficient and isotropic stencils \cite{7161520}. YASK \cite{yount2016yask} could improve data reuse by using common expression elimination and unrolling based on their vector-folding methods with fine-grained blocks \cite{yount2015vector}, which is less feasible for high-order complex stencils \cite{zhao2019exploiting}. 
Zhao \cite{zhao2019exploiting} designs a greedy algorithm to decide the part of the computation with high reuse. 
For other parts not identified by the algorithm, they still utilize the original computation by gather operations. Rawat \cite{rawat2018register} utilizes a DAG of trees with shared leaves to describe the stencil computation, and devises a scheduling algorithm to minimize register usage by reordering instructions on GPUs. 
Common subexpression elimination (CSE) \cite{aho1976code} is presented to reduce the redundant computation in successive iterations of the same loop by reusing partial sums of a subexpression. This method relies heavily on loop unrolling to find specific expressions.
Deitz extends the CSE method as Array Subexpression Elimination (ASE) \cite{deitz2001eliminating} by creating an abstraction called a neighborhood tablet. 
Since the ASE reuses partial sums by subtablets via temporary variables, scalar dependences are newly introduced and hinder the instruction-level parallelization by compilers.  
  
Tiling \cite{Irigoin.Triolet:popl88,McKeller.Coffman:cacm69,Lam+:asplos91,Wolf.Lam:pldi91,Wolfe:sc89}
is one of the most powerful transformation techniques
to explore the data locality of multiple loop nests.
Notably work for stencil computations includes hyper-rectangle tiling \cite{Ding.He:sc01,Rastello.Dauxois:ipdps02,Rivera.Tseng:sc00,Nguyen+:sc10},
time skewed tiling \cite{Song.Li:pldi99,Wonnacott:ijpp02,Jin+:sc01},
diamond tiling  \cite{bondhugula2008practical,bandishti2012tiling},
cache oblivious Tiling  \cite{10.1145/1989493.1989508,strzodka2010cache,Frigo.Strumpen:ics05}, 
split-tiling \cite{10.1145/2464996.2467268}
and tessellating \cite{yuan2019tessellating}.
Wonnacott and Strout present a comparison on the scalability of many existing tiling schemes \cite{Wonnacott.Strout:impact13}.
Most of these techniques are compiler transformation techniques
and this paper integrated the new proposed layout with the tessellation scheme for simplifying the implementation.
For stencil computations, a variety of auto-tuning frameworks \cite{christen2011patus,gysi2015modesto,kamil2010auto,zhang2012auto} have been presented by using varied hyper-rectangular tiles to exploit data reuse alone. However, redundant computations are involved in these work to resolve the introduced inter-tile dependencies that hinder the concurrent execution of shaped tiles on different cores.  

\section{\label{sec:conclusion}Conclusion}

In this paper, we propose a novel transpose layout to overcome the data alignment conflicts efficiently \black{for vectorization in the data space. Then a computation folding approach by reducing the redundancy of arithmetic calculations in time iteration space is devised }on the basis of the proposed transpose layout. 
Furthermore, we describe how the proposed vectorization scheme is optimized with shifts reusing for enhancing data reuse and integrated with tessellate tiling for improving data locality. With the qualitative analysis and quantitative experiments, we demonstrate that significant performance improvements are achieved by our vectorization scheme over state-of-the-art products such as Intel's ICC and recent work \cite{10.1145/2464996.2467268,10.1145/3126908.3126920}. 


\section{Acknowledgements} 
This work is supported by the National Key R\&D Program of China under Grant No.2016YFB0200803.

\bibliographystyle{ACM-Reference-Format}
\bibliography{sample-base}

\end{document}